\documentclass[a4paper,10pt,twoside]{cpc-hepnp}
\usepackage{upgreek,fancyhdr}
\usepackage{hyperref}
\usepackage{graphicx}
\usepackage{float}
\usepackage{stfloats}
\usepackage{multicol}
\usepackage{booktabs}
\usepackage{amssymb,bm,mathrsfs,bbm,amscd}
\usepackage[tbtags]{amsmath}
\usepackage{lastpage}
\usepackage{color,xcolor}
\usepackage{dcolumn}
\usepackage{tabularx,multirow,blindtext}
\usepackage[font=small]{caption}
\usepackage{setspace}

\newenvironment{figurehere}
{\def\@captype{figure}}
{}

\begin{document}

\fancyfoot[C]{\small \thepage}

\title{Charge-changing cross section measurements of 300 MeV/nucleon $^{28}$Si on carbon and data analysis}

 \author{Chang-Jian Wang$^{1}$
 \quad Ge Guo$^{1}$
 \quad Hooi Jin Ong$^{2}$
 \quad Yu-Nan Song$^{2,3}$
 \quad \\Bao-Hua Sun$^{1;1)}$\email{bhsun@buaa.edu.cn}
 \quad Isao Tanihata$^{1}$
 \quad Satoru Terashima$^{1}$
 \quad Xiu-Lin Wei$^{1}$
 \quad \\Jun-Yao Xu$^{1}$
 \quad Xiao-Dong Xu$^{2;2)}$\email{xiaodong.xu@impcas.ac.cn}
 \quad Ji-Chao Zhang$^{1}$
 \quad Yong Zheng$^{2;3)}$\email{zhengyong@impcas.ac.cn}
 \quad \\Li-Hua Zhu$^{1}$
\quad Yong Cao$^{1}$
\quad Guang-Wei Fan$^{4}$
\quad Bing-Shui Gao$^{2}$
\quad \\Jia-Xing Han$^{1}$
\quad Guang-Shuai Li$^{1}$
\quad Chen-Gui Lu$^{2}$
\quad Hao-Tian Qi$^{1}$
\quad \\Yun Qin$^{1}$
\quad Zhi-Yu Sun$^{2}$
\quad Lu-Ping Wan$^{1}$
\quad Kai-Long Wang$^{2}$     
\quad \\Shi-Tao Wang$^{2}$
\quad Xin-Xu Wang$^{1}$    
\quad Mei-Xue Zhang$^{1}$
\quad Wen-Wen Zhang$^{1}$
\quad \\Xiao-Bin Zhang$^{1}$
\quad Xue-Heng Zhang$^{2}$
\quad Zi-Cheng Zhou$^{1}$
}

\maketitle
 
\address{
	$^1$ School of Physics, Beihang University, Beijing 100191, China\\
	$^2$ Institute of Modern Physics, Chinese Academy of Sciences, Lanzhou, 730000, China\\
    $^3$ School of Nuclear Science and Technology, University of Chinese Academy of Sciences, Beijing, 100049, China\\
    $^4$ School of Chemical Engineering, Anhui University of Science and Technology, Huainan 232001, China\\
}

\begin{abstract}
Charge-changing cross section ($\sigma_{\text{cc}}$) measurements via the transmission method have made important progress recently aiming to determine the charge radii of exotic nuclei. In this work, we report a new $\sigma_{\text{cc}}$ measurement of 304(9) MeV/nucleon $^{28}$Si on carbon at the second Radioactive Ion Beam Line in Lanzhou (RIBLL2) and describe the data analysis procedure in detail. This procedure is essential to evaluate the systematic uncertainty in the transmission method.
The determined $\sigma_{\mathrm{cc}}$ of 1125(11) mb is found to be consistent with the existing data at similar energies. The present work will serve as a reference in the $\sigma_{\text{cc}}$ determinations at RIBLL2.

\end{abstract}


\begin{multicols}{2}

\section{Introduction} \label{sec_intro}

The charge-changing cross section ($\sigma_{\text{cc}}$) is defined as the total probability of removing at least one proton from the projectile nucleus in the collision with the target nucleus. By combining with the Glauber-type model calculations, the precise $\sigma_{\text{cc}}$ data have been used to deduce the root-mean-square ($rms$) point-proton radius of atomic nuclei~\cite{Yamag2011PRL,Teras2014PEPT,Estra2014PRL,Ozawa2014PRC,Kanun2016PRL,Bag2019PLB,Kaur2022PRL,Tanak2022PRC,Tran2016PRC}. This approach can be in principle applied to any nuclei that can be produced in flight and thus provide a complementary method to the electron scattering and the isotopic shift method. Moreover, 
studying the charge-changing reactions of mirror nuclei has been proposed as a new method to constrain the density dependence of the symmetry energy~\cite{Xujun2022PLB}.

Experimentally, $\sigma_{\text{cc}}$ is determined by the transmission method, which has been applied well to the interaction cross section measurements.
$\sigma_{\text{cc}}$ is calculated as follows:
\begin{equation}\label{eq1}
  \sigma_{\text{cc}} = -\frac{1}{t}\ln(\frac{N_{\text{out}}}{N_{\text{in}}}) \;,
\end{equation}
where \textit{t} is the target thickness. $N_{\text{in}}$ and $N_{\text{out}}$ are the number of incident ions and the number of outgoing ions with the proton number ($Z$) no less than that of the incident nucleus, respectively. In this method, it is essential to identify and count precisely both the incident and non-reacted ions. To eliminate the effect of reactions in materials other than the target, measurements without reaction target (so-called empty target) but with a match of incident beam energy should be conducted. Therefore, Eq.~\ref{eq1} is then written as:
\begin{equation}
  \sigma_{\text{cc}} = -\frac{1}{t}\ln(\frac{\textit{$\Gamma$}}{\textit{$\Gamma$}_0}) \;,
\end{equation}
where $\textit{$\Gamma$}=N_{\text{out}}/N_{\text{in}}$. $\textit{$\Gamma$}$ and $\textit{$\Gamma$}_0$ are the ratios with and without the reaction target, respectively.

At the Heavy Ion Research Facility in Lanzhou (HIRFL)~\cite{xiajw2016csb,ZhanW2010NPA}, a series of experiments have been performed using the second Radioactive Ion Beam Line in Lanzhou (RIBLL2)~\cite{SUN201878}, including the $\sigma_{\text{cc}}$ measurements for $psd$-shell nuclei at around 300 MeV/nucleon~\cite{Zhao2019NN,Zhaojw2020NPR,Sun2021CSB,Liguangshuai2023PRC}, the projectile fragmentation cross section of 260 MeV/nucleon $^{18}$O  on carbon~\cite{Xuxd2022CPC}, and the single-nucleon removal cross sections of $^{14-16}$C and $^{14}$O on carbon~\cite{Sunzy2014PRC,Zhaoy2019PRC,Sunyz2021PRC}. Recently, we extended the $\sigma_{\text{cc}}$ measurements for neutron-rich $sd$-shell nuclei to the F4 focal plane area of RIBLL2, where an experimental platform (namely RIBLL2-F4) has been completed.

Based on the data obtained from the experiment at RIBLL2-F4, the present work aims to provide a detailed description of the data processing procedures for $\sigma_{\text{cc}}$ measurements by employing the transmission-type technique. Such details are often missing in literature but play a crucial role in understanding the discrepancies among the data obtained at similar energies but in different experiments. Moreover, we are aware that there is also a different definition of $\sigma_{\text{cc}}$ in literature, in which $N_{\text{out}}$ refers to the total number of outgoing ions with $Z$ unchanged. In comparison with our definition, the key is whether to count those outgoing ions with $Z$ increased after the collision which should originate from the charge-exchange process. Although the total event number for $Z$-increased is generally very small, it might affect the final cross section in case of hydrogen target where the ratio between charge-exchange reaction cross sections and charge-changing cross sections is large. 

The present paper is organized as follows. An overview of the experiment is given in Sec.~\ref{sec_exp}. Then the details of data analysis procedures are introduced in Sec.~\ref{sec_ana}. In Sec.~\ref{sec_discuss}, we report the $\sigma_{\rm cc}$ obtained for the $^{28}$Si on a natural carbon target at 304(9) MeV/nucleon. Finally, a brief summary is given in Sec.~\ref{sec_sum}.

\section{Experiment}\label{sec_exp}
 \begin{figurehere}
  \centering
  \includegraphics[width=3in]{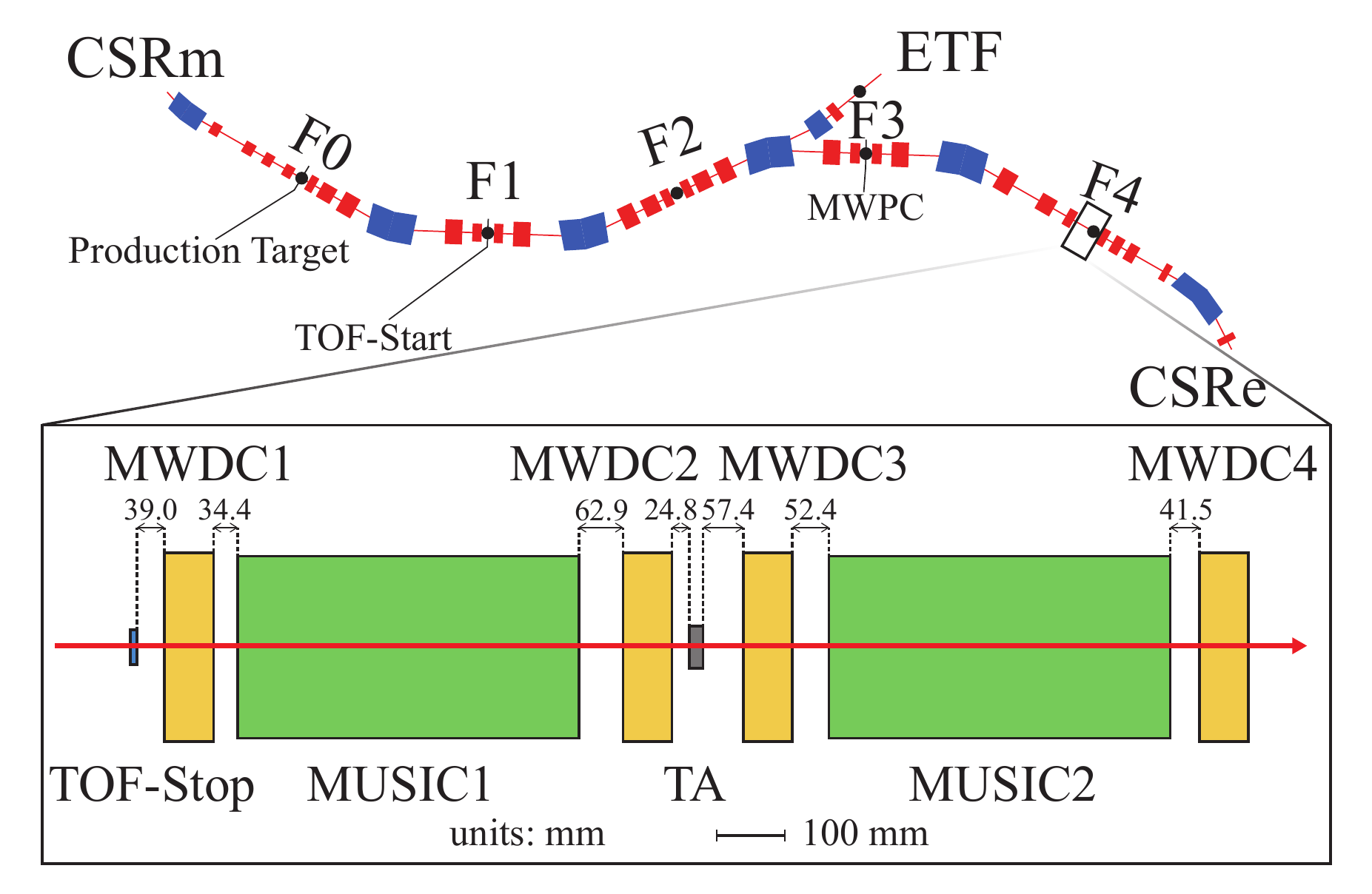}
      \figcaption{(Color online) Layout of RIBLL2 and scheme of detector setup for the charge-changing cross section measurements in top view (in scale). The dimensions are proportional to the real sizes. The CSRm refers to the main cooler storage ring. F1, F2, F3, and F4 represent focal planes of RIBLL2. The ETF indicates an experimental area called External Target Facility and the CSRe is the experimental cooler storage ring. The distances between neighboring devices are labeled.}
  \label{fig:ribll2}
\end{figurehere}

\begin{figure*}
    \centering
    \includegraphics[width=7in]{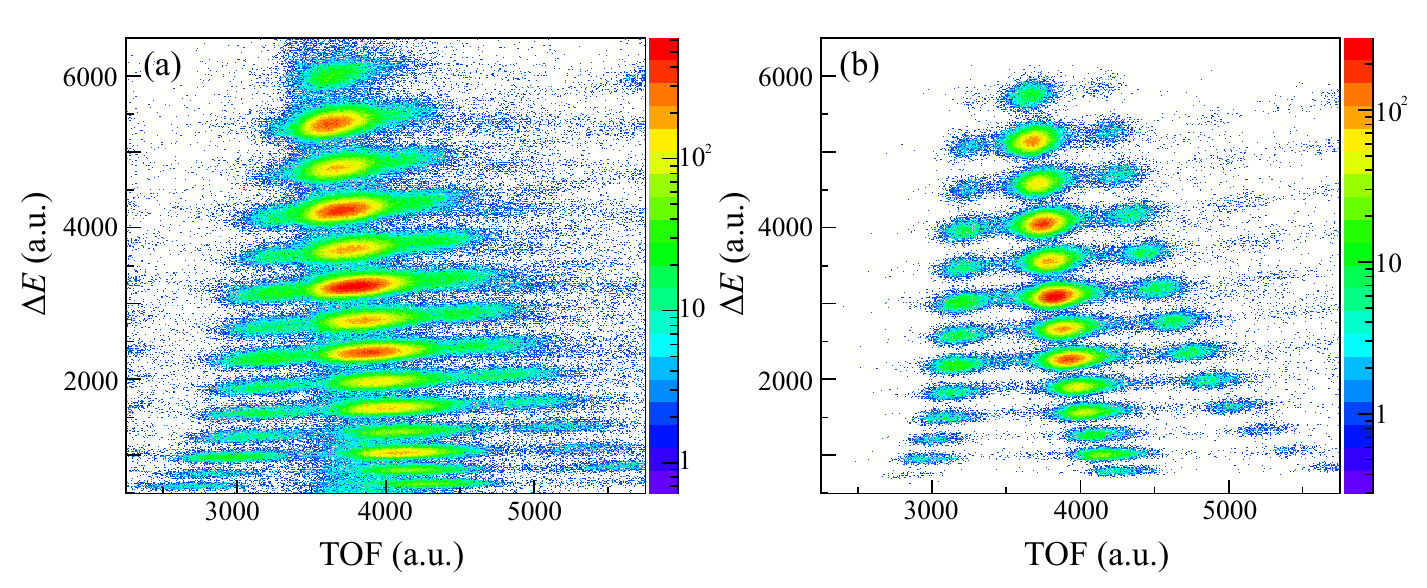}
	\figcaption{(Color online) Particle identification of cocktail beams produced via $^{40}$Ar + $^9$Be at 400 MeV/nucleon. 
	(a) Raw $\Delta E$-TOF spectrum. (b) Same as (a) but after data processing.}
    \label{fig:pid}
\end{figure*}
In the experiment, a primary $^{40}$Ar beam was accelerated to 400 MeV/nucleon at the main cooler storage ring (CSRm) and then guided in slow extraction mode to impinge on a 10-mm thick beryllium target located at the F0 focus of the RIBLL2~\cite{SUN201878}. Tens of isotopes were produced via fragmentation reactions. The secondary beam of interest was then transported by the RIBLL2 to bombard the reaction target located at the final focus (F4) of RIBLL2.

The layout of RIBLL2 and the detector system used in the experiment are sketched in Fig.~\ref{fig:ribll2}. At the first focus F1, a scintillator detector was positioned to give the Time-of-Flight (TOF) start (TOF-start) information. A multi-wire proportional chamber (MWPC) with an intrinsic position resolution 199.9 $\mu$m ($\sigma$) in X direction and 201.2 $\mu$m ($\sigma$) in Y direction was installed at F3~\cite{ZhangX2017NPR}.
The reaction target (TA), 
of natural carbon with the thickness of 1.855 g/cm$^2$, was placed at the center of the final focus F4, where the cross section measurements were performed. 

Upstream of the reaction target, a fast timing scintillator detector (TOF-stop)~\cite{Zhao2016NIMA,linwj2017CPC}, a multiple sampling ionization chambers (MUSIC1)~\cite{ZHANGXH2015NIMA,Zhao2019NIMA} sandwiched by two multi-wire drift chambers (MWDC1 and MWDC2) were installed. 
The TOF-start and TOF-stop detectors were read out by both ends. The time intervals between the left-side readout of TOF-start and TOF-stop, right-side readout of TOF-start and TOF-stop, and both sides of TOF-stop are generated by a Time-to-Amplitude converter (TAC). The active sizes of the TOF-start and TOF-stop detectors are 100$\times$100$\times$1 mm$^3$ and 50$\times$50$\times$1 mm$^3$, respectively. Downstream of the reaction target, a similar setup consisting of one ionization chamber (MUSIC2) and two drift chambers (MWDC3 and MWDC4) were placed, as shown in Fig.~\ref{fig:ribll2}. Both MUSIC1 and MUSIC2 have the same structure with four cells and each cell provides an individual energy deposition ($\Delta E$) signal. Their active area and length are $\Phi$90 mm and 488 mm, respectively.  
All the four MWDCs have the same configuration with an effective volume of 160$\times$160$\times$70 mm$^3$ and can reach a position with the resolution of about 100 $\mu$m ($\sigma$)~\cite{Zhaojw2020NPR}. The combination of two MWDCs before and after TA was utilized to determine the trajectories of incoming and outgoing particles.

The coincidence between TOF-start and TOF-stop was employed as the trigger of the events.
The TOF and energy-loss ($\Delta E$) for incident ion were determined on an event-by-event basis by the TOF-start and TOF-stop scintillator detectors, and by the MUSIC1, respectively.
The magnetic rigidity ($B\rho$) was set to deliver $^{28}$Si$^{14+}$ in the central orbital of RIBLL2.

 The particle identification (PID) for the incoming ions upstream of the reaction target was achieved by employing the standard $\Delta E$-TOF-$B\rho$ method. Fig.~\ref{fig:pid}(a) shows the raw PID spectrum for the $^{28}$Si setting without any data filtering and processing. Although dozens of ion species can be in principle identified, it is getting difficult to distinguish the isotopes with larger $\Delta E$ (thus high $Z$). As a comparison, Fig.~\ref{fig:pid}(b) shows the final particle identification (PID) after data processing, in which the neighboring isotopes are well separated. The details of data processing will be described in the following section. 

\section{Data analysis}\label{sec_ana}

In this section, we introduce the data analysis procedures for the determination of charge-changing cross section step by step.  

\subsection{Detector response}

As the first step of data processing, one needs to ensure all the detectors have good and correct responses. Incorrect or improper responses, e.g., due to limited detection efficiency, overload, pileup, and electronic errors, have to be examined and excluded from the data.

Pileup refers to the case that two or more successive incident ions are recorded as a ``fake" single event. This often occurs at high beam rates for the detector (e.g., MUSIC) with a slow time response. It is worth mentioning that pileup is a crucial issue when the beam is delivered from CSRm in a slow extraction mode. Micro-structures of the beam have been observed in the order of 500 ns, 1 ms, and 20 ms. In our experiment, we have set up an electronic circuit for pileup-event rejection with the fast TOF detectors. Moreover, we employed the multihit-TDC for time digitizing, which allows us to record all the successive events in a time interval of 12 $\mu$s before and 8 $\mu$s after the trigger. This also helps to deal with the pileup events.

\begin{figurehere}
    \centering
    \includegraphics[width=3in]{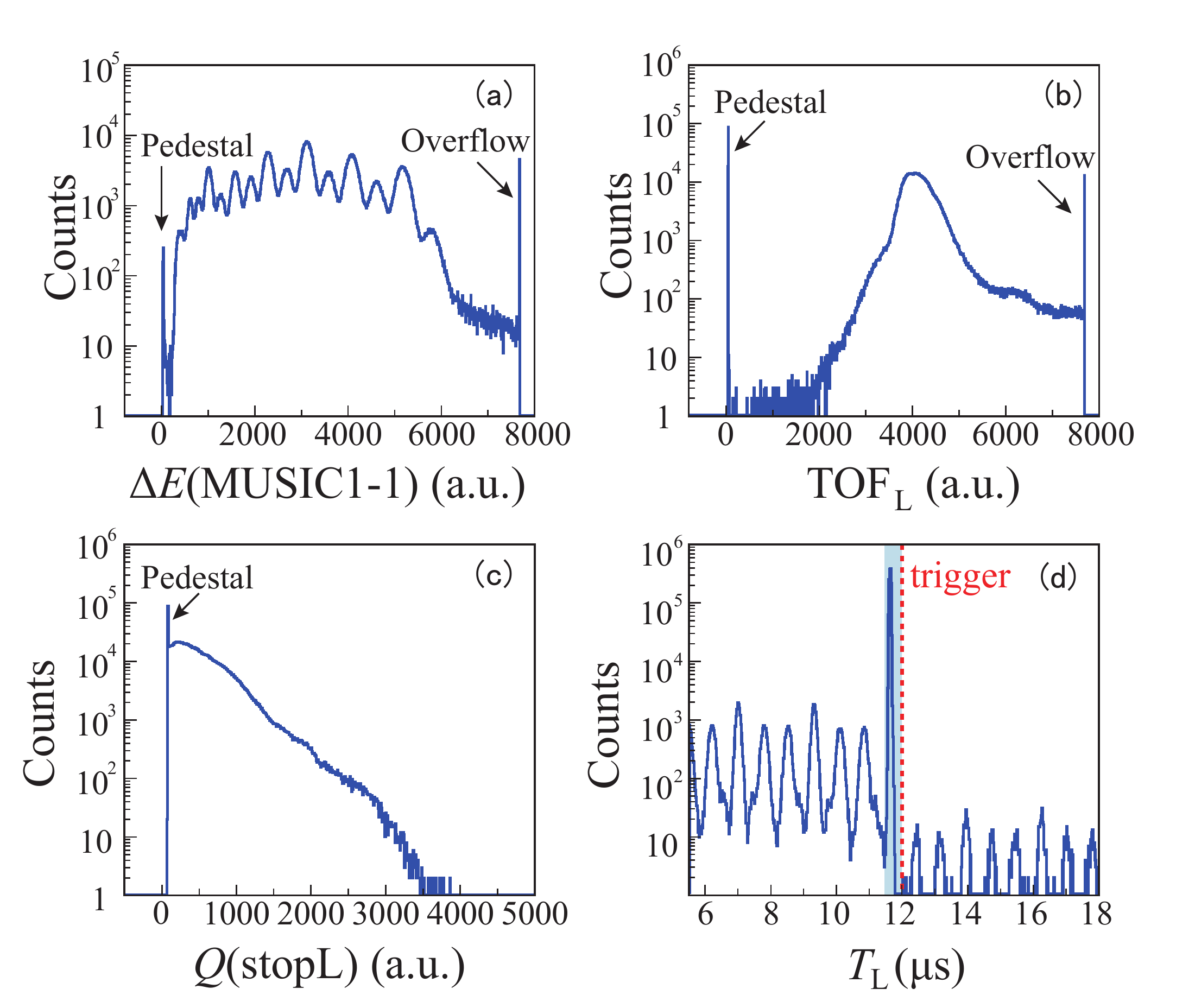}
	\figcaption{(Color online) Typical spectrum recorded by MUSIC1, TOF and MWDC detectors. Overflow and pedestal events are indicated. (a) Energy loss ($\Delta E({\rm MUSIC1\mbox{-}1})$) spectrum measured by the first cell of MUSIC1. (b) Time difference (TOF$_{\rm L}$) between the left-side readout of TOF-start and TOF-stop. (c) Charge signal from the left-side readout of TOF-stop, $Q({\rm stopL})$. (d) Registered timing distribution of the first hit on MWPC for each event, $T_{\rm L}$. }
    \label{fig:basic}
\end{figurehere}

Besides the pileup, the pedestal and overflow signals should be also checked. Overflow indicates that the digitized values exceed the acceptance range of electronics, while the pedestal corresponds to the electronics noise when no event is recorded and/or the under scale events. One should examine each detector to exclude both overflow and pedestal signals. As examples, Fig.\ref{fig:basic} (a,b,c) depicts the energy loss distribution from the first cell of MUSIC1 (MUSIC1$\mbox{-}$1), the left-side readout of the time difference between TOF-start and TOF-stop, the left-side readout of charge information from TOF-stop, respectively. One can see the pedestal at near zero and the overflow with sharp peaks. Events with signals in these overflow and pedestal peaks are eliminated. The same operation has been applied to all the detectors along the beamline before the reaction target.

The positions determined by MWDCs are crucial to deduce the ion's trajectory thus to evaluate the acceptance of detectors. 
In MWDCs, each layer of wires gives two possible positions for one correct response.
The trajectory upstream of the target is computed 
by finding the linear fit which can describe best the combinations of hit positions. The hit positions were deduced by the eight layers of wires in 2 MWDCs, four layers for $X$ direction and four layers for $Y$ direction. 
We exclude those events whose linear fit has a sum of squares error (SSE) of more than 0.04 in $X$ and $Y$ directions.
At F3, the hit position of the particle at MWPC is determined by two layers of wires. Taking the left-side readout from MWPC as an example, we show the registered timing distribution of the first hit $T_{\rm L}$ for each event in Fig.~\ref{fig:basic}(d). One can see the micro-structure of the beam in a period of about 1.5 $\mu$s. For our purpose, we choose the events with the first hit occurring only within 450 ns before the trigger, as illustrated by the shadowed area. The same operation has been applied to the right-side readout of MWPC.

\subsection{Consistency check}

In the experiment, we fully utilized the potential of the detector setup to record as much information as possible. For instance, both the timing and the energy loss information induced by the ion on TOF-stop detector were recorded. Therefore, we can check the consistency of signals not only from the same detector but also from different detectors in coincidence. 

The $\Delta E$ in MUSIC is far small than the total kinetic energy of ions, and the values measured by four different cells of MUSIC should be consistent with each other. Fig.~\ref{fig:correlation}(a) displays the $\Delta E$ measured by the first and second cells of MUSIC1. One can clearly see that the data at the first two cells, $\Delta E({\rm MUSIC1\mbox{-}1})$ and $\Delta E({\rm MUSIC1\mbox{-}2})$, are consistent for those events inside the red polygon. Furthermore, we checked the correlation between the energy loss measured by MUSIC1, $\Delta E_{\rm 1}$, and TOF-stop, $\Delta E({\rm TOF\mbox{-}stop})$. The results are shown in Fig.~\ref{fig:correlation}(b). Here the $\Delta E_{\rm 1}$ is defined as the average value of the energy deposition measured by the four cells of MUSIC1, while the $\Delta E({\rm TOF\mbox{-}stop})$ was evaluated as the geometrical mean of the integrated charges from the both-side readouts~\cite{linwj2017CPC}. Similar to Fig.~\ref{fig:correlation}(a), those events lying within the dashed-line area in Fig.~\ref{fig:correlation}(b) demonstrate the consistency. 

\begin{figurehere}
	\centering
	\includegraphics[width=3in]{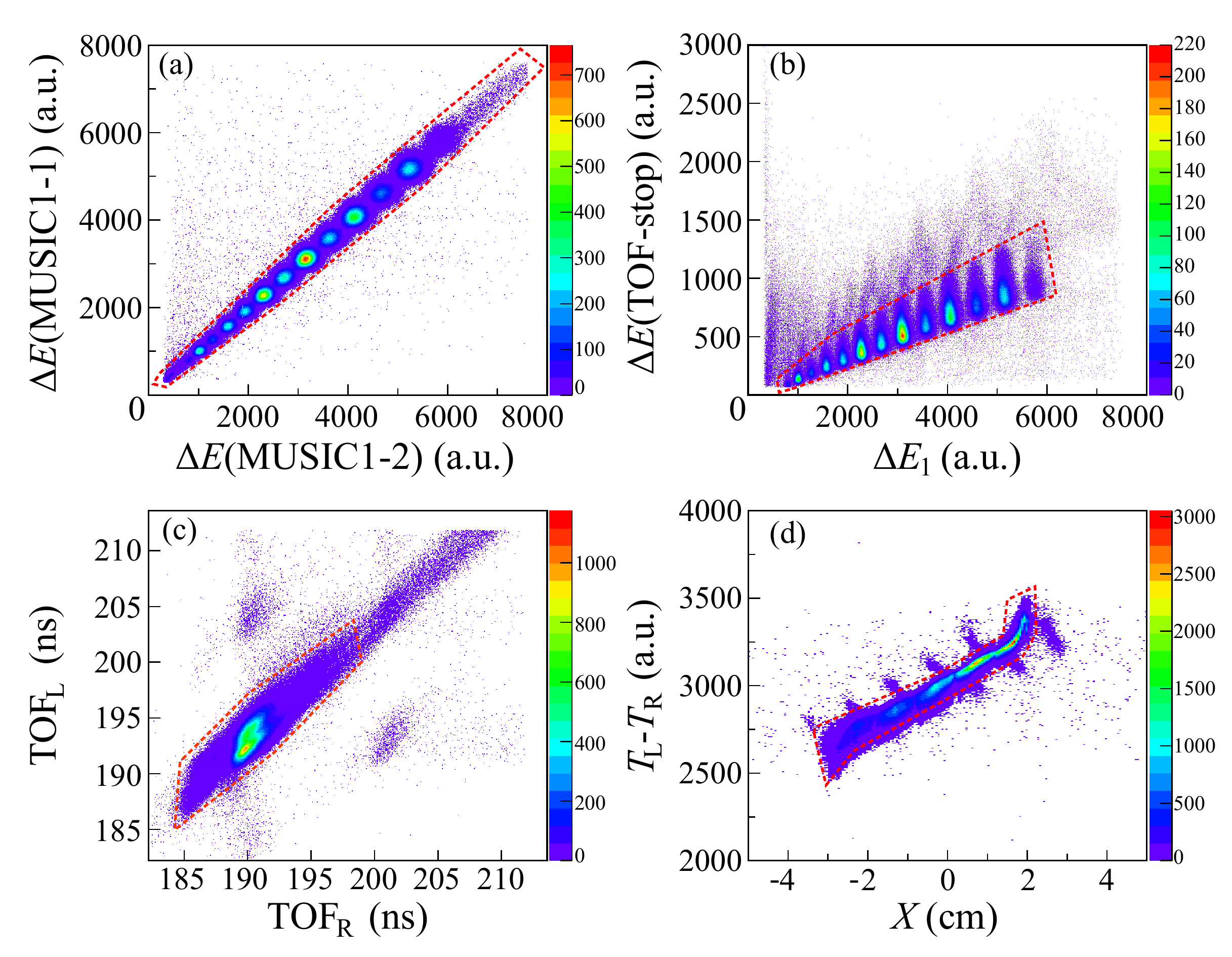}
	\figcaption{(Color online) Correlations in the various detector responses. (a) Energy deposition in the first cell of MUSIC1 versus that in the second cell of MUSIC1. (b) Energy deposition in TOF-stop versus that in MUSIC1. (c) TOF recorded by the left-side and right-side readouts of the TOF-start and TOF-stop detectors. (d) Ion positions deduced by the two-side time difference of the TOF-stop detector and those measured by MWDCs. The dashed lines define the events with consistent responses in detectors along the beam line.}
	\label{fig:correlation}
\end{figurehere}

TOF can be determined individually by the left-side (TOF$_{\rm L}$)  and right-side (TOF$_{\rm R}$) readout of TOF-start and TOF-stop detectors. A good coincidence between them will guarantee a good TOF measurement as indicated in Fig.~\ref{fig:correlation}(c). 
TOF$_{\rm L}$ or TOF$_{\rm R}$ of more than 200 ns (the upper right region in Fig.~\ref{fig:correlation}(c))  has been found to be correlated with the electronic reflection signals and the relevant events are excluded from further analysis.

The position distributions of ions at TOF-stop can be deduced from MWDCs and also the time difference between the both-end readouts of TOF-stop. 
Fig.~\ref{fig:correlation}(d) shows the positions determined by these two methods. It is evident that the events lying within the red-lined area exhibit a good correlation and are chosen for further analysis.

\subsection{Positions at the TOF-stop detector and target}

After the detector response and consistency check, one should examine the incident ion's positions at TOF-stop detector and target. Fig.~\ref{fig:stop} displays the position distribution of all ions at the TOF-stop detector. One can see that most of the events are inside a square area with a clear boundary, which indicates the edge of the TOF-stop detector. Actually, the active area of TOF-stop is 5.0 $\times$ 5.0 cm$^2$. Given the fact that the particles that hit the detector edge may have a chance of being recorded as abnormal signals, we applied a rectangular gate (see the red rectangle in Fig.~\ref{fig:stop}) with a dimension of 4.8$\times$4.8 cm$^2$ to select events. 

\begin{figurehere}
    \centering
    \includegraphics[width=3in]{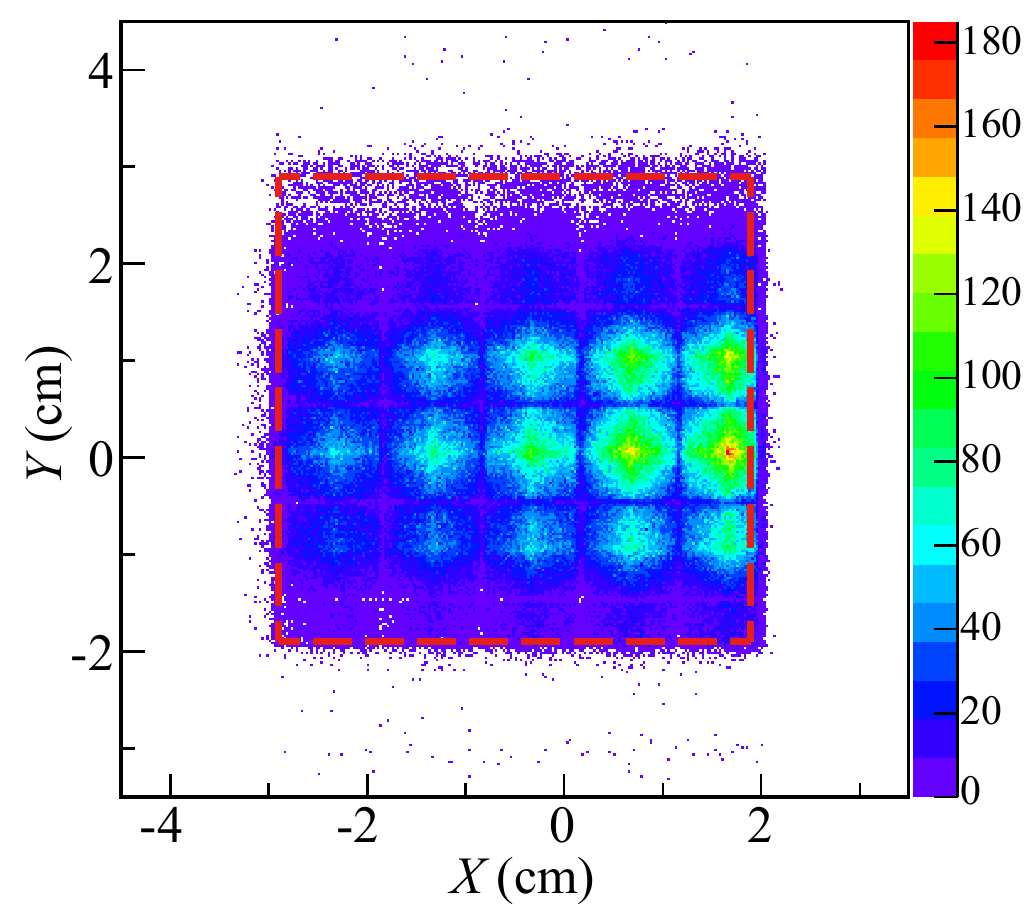}
    \figcaption{(Color online) Ion's position distribution at the TOF-stop detector. The size of the TOF-stop detector is 50 mm $\times$ 50 mm. The red rectangle has a dimension of 48 mm $\times$ 48 mm.}
    \label{fig:stop}
\end{figurehere}

\begin{figurehere}
    \centering
    \includegraphics[width=3in]{targetphoto.pdf}
    \figcaption{(Color online) Photograph and engineering drawing of the carbon target and the aluminum support frame. The radii of the target and the support frame are 30 mm and 25 mm, respectively.}
    \label{fig:engineer}
\end{figurehere}

The left part of Fig.~\ref{fig:engineer} is a photo of the target system including the carbon target and the aluminum support frame. The corresponding engineering drawings are shown in the middle and right part of Fig.~\ref{fig:engineer}. 
In such a structure, ions can pass through three possible target thicknesses, i.e., the carbon target, both the carbon target and the aluminum frame, and the aluminum frame only, depending on their trajectories. Ions that travel through thicker matter would lose more energy than anticipated and vice versa. This will result in different energy depositions in MUSIC2. Such a difference can be better highlighted by comparing $\Delta E_{\rm 1}$ and the average energy deposition in four cells of MUSIC2, $\Delta E_{\rm 2}$, which are illustrated in Fig.~\ref{fig:targetcircle}(a). The events circled by the red (deep blue) dashed lines, representing the higher (lower) energy depositions than the central ones, are labeled as Region-A (Region-B). The position distributions at TA for the events in Region-A and Region-B are further shown in Fig.~\ref{fig:targetcircle}(b) and (c), respectively. Within the ring-like distribution in Fig.~\ref{fig:targetcircle}(b), the relevant ions penetrate through both the carbon target and the aluminum frame. The center of  the target frame can be determined by this ring precisely. Within the right part of Fig.~\ref{fig:targetcircle}(c), particles transport only through the aluminum frame. Fig.~\ref{fig:targetcircle}(d) shows the $\Delta E_{\rm 2}$ versus the distance from the target center to the hit position ($R$). The systematic shift in $\Delta E_{\text{2}}$ at around 2.6 cm is well correlated with the size of the frame radius of 2.5 cm. 
This demonstrates the current position determination is robust and consistent. Only these ions hitting on the carbon target will be considered. 

\begin{figurehere}
    \centering
    \includegraphics[width=3in]{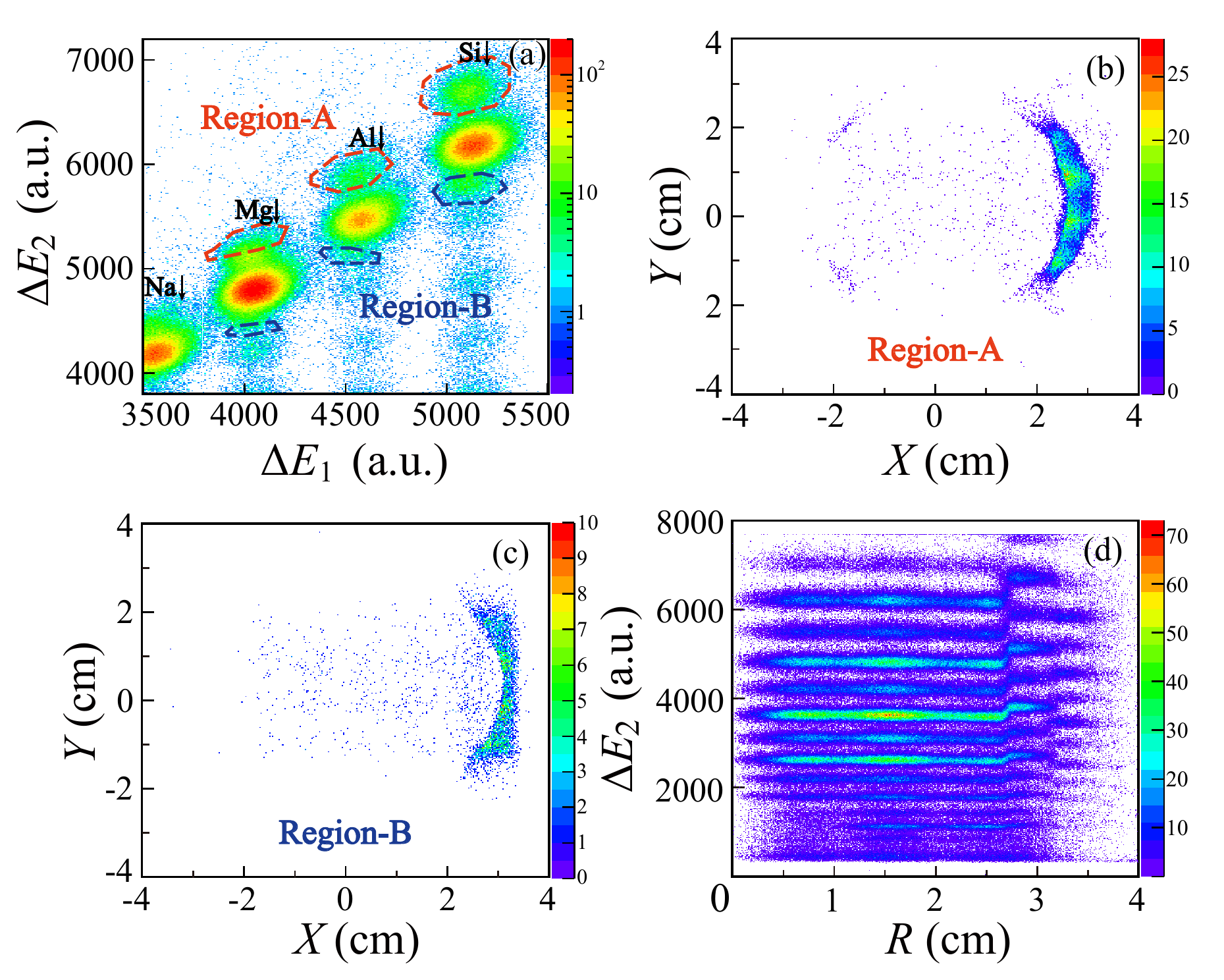}
    \figcaption{(Color online) Procedure to refine the incident position on the target. 
	(a) Energy loss in MUSIC2 versus that in MUSIC1. Events with higher and lower $\Delta E_{\rm 2}$ were labeled by the red dotted line (Region-A) and blue dotted line (Region-B), respectively. (b) Position distribution at TA for the events at Region-A. (c) Position distribution at TA for the events at Region-B. (d) $\Delta E_{\rm 2}$ as a function of the distance from the hit point to the target center. }
    \label{fig:targetcircle}
\end{figurehere}

\subsection{TOF correlation with position}

In the secondary beam experiments, it is known that only those ions with the right magnetic rigidity ($B\rho_0$) will travel in the central orbit of the beam line.
At RIBLL2, F2 and F4 are the achromatic focal planes, while F1 and F3 are the dispersive planes.  
At the dispersive foci, the precise $B\rho$ values vary with the $X_F$ positions as~\cite{Fangf2021NPR}:
\begin{equation}
    B = B \rho_0 (1+\frac{X_F}{(X|\delta)}) \;,
\end{equation}
where $(X|\delta)$ denote the momentum dispersion. Ions with a different $B\rho$ will be guided to different $X$ positions at the focal plane. Therefore, the measured $X$ positions at the F3 and F4 foci in our experiment can be employed to improve the resolving ability of the particle identification~\cite{Fangf2021NPR} by the TOF-$X$ correlations. 

It is necessary to select the nuclei of interest approximately in the raw TOF-$\Delta E$ graph. For our purpose, the $^{28}$Si ions are selected. Firstly, the dependence of TOF on $X$ position at F3 is examined to follow a linear trend, as shown by Fig.~\ref{fig:revise} (a). The spectrum after the linear correction is shown in Fig.~\ref{fig:revise} (b). Similarly, the dependence of the corrected TOF on the $X$ position at F4 can be described by a second-order polynomial, as depicted by the red curve in Fig.~\ref{fig:revise} (c). Finally, we correct the TOF again by using the fit curve and show in Fig.~\ref{fig:revise} (d). After the two-step corrections, the final PID plot with improved resolving power before the reaction target is shown in Fig.~\ref{fig:pid} (b).

\begin{figurehere}
    \centering
    \includegraphics[width=3in]{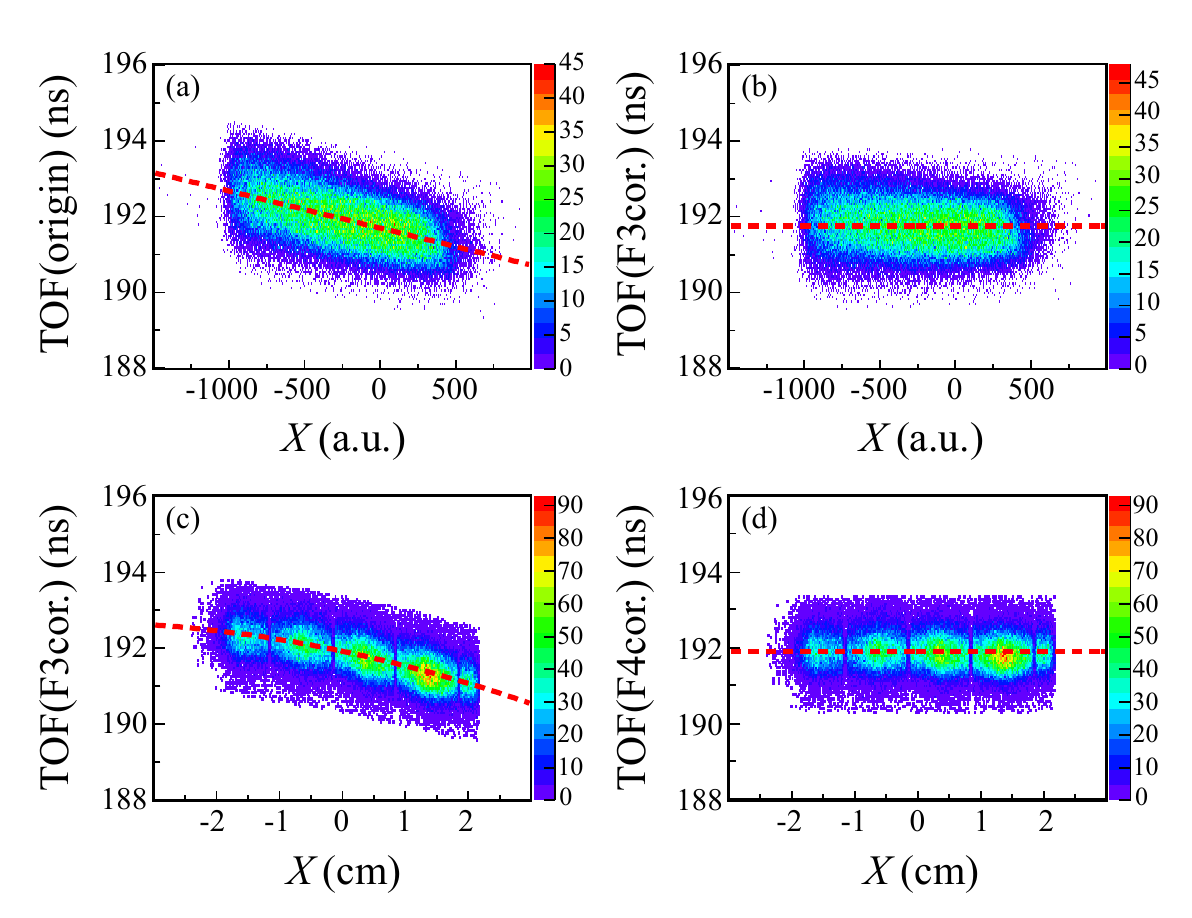}
    \figcaption{(Color online) (a) The original TOF for the $^{28}$Si ions as a function of $X$ position at F3. The linear fit to the spectrum is shown by the dashed lines. (b) Same as (a) but after the correction of the dependence of TOF on $X$ at F3. (c) The corrected TOF (TOF(${\rm F3cor.}$)) from panel (b) versus $X$ positions at TA. (d) Same as (c) but after the correction of the dependence of TOF(${\rm F3cor.}$) on $X$ at F4.
    }
    \label{fig:revise}
\end{figurehere}

\subsection{Acceptance insurance}

 \begin{figurehere}
	\centering
	\includegraphics[width=3in]{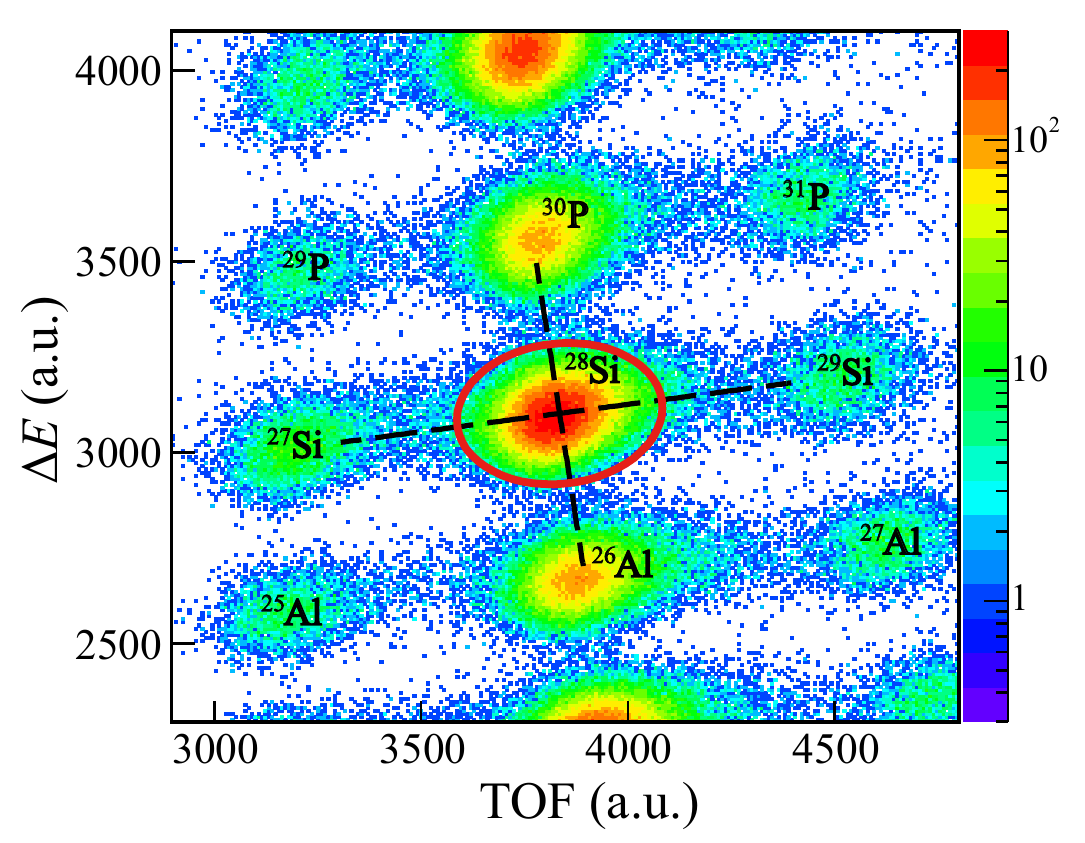}
	\figcaption{(Color online) Particle identification around $^{28}$Si.}
	\label{fig:pidnin}
\end{figurehere} 

In Fig.~\ref{fig:pidnin}, we present the particle identification plot for the ions around $^{28}$Si. The red ellipse cut is employed to select $^{28}$Si. We carefully analyzed the possible contamination from the neighboring isotopes and found that the contamination is less than 0.01$\%$. 

\begin{figurehere}
    \centering
    \includegraphics[width=3in]{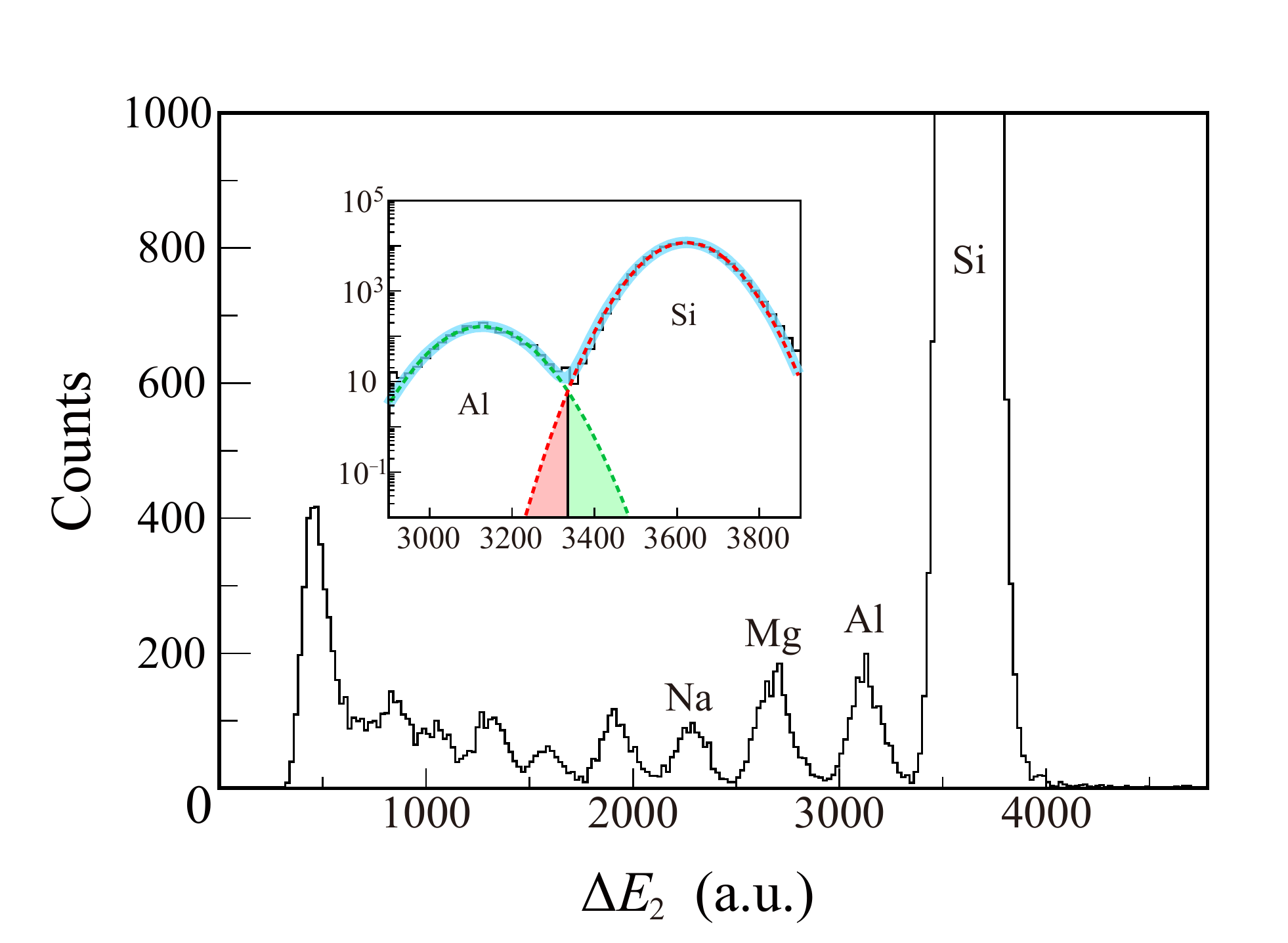}
    \figcaption{(Color online) Energy deposition of the outgoing ions after reaction target at MUSIC2 relevant to the incident $^{28}$Si. Insert is the zoom-in. The blue line is the two-Gaussian fit for the Si and Al peaks. The green and red dashed lines are the deconvoluted functions for the Al and Si peaks, respectively. The red and green shadowed areas are parts of the deconvoluted functions. }
    \label{fig:music2gauss}
\end{figurehere}

After selecting the $^{28}$Si ions before the target, we examine the energy deposition in MUSIC2 of outgoing ions, as displayed in Fig.~\ref{fig:music2gauss}. The inset of Fig.~\ref{fig:music2gauss} depicts a zoom-in of the Al peak and Si peak. To deconvolute the spectrum, we use a two-Gaussian function to fit the total Energy deposition in MUSIC2 ($\Delta E_{\rm 2}$) spectrum of Al-Si peaks. The relevant Gaussian fits are illustrated by the green and red curves, respectively. 
The green and red shadowed areas are used to estimate the number of events that belong to Al isotopes but are included in the Si peak, and the number of events that belong to Si isotopes but are included in the Al peak, respectively. The relevant event counts are $N_\text{Al-Si}$ and $N_\text{Si-Al}$, respectively.
For the charge-changing cross section, $N_{\rm out}$ is the total event number of outgoing particles with $Z$ no less than 14. It is computed as $N_\text{Si}-N_\text{Al-Si}+N_\text{Si-Al}$, where $N_\text{Si}$ refers to the count of events with $\Delta E_2$ larger than intersection of the Al-Si peaks. The leakage of events from the Si peak to the Al peak and vice versa has been corrected. The difference between the two Gaussian fits and the experimental data in the range of interest, together with the statistic uncertainty, have been taken as the uncertainty of $N_\text{out}$.  


As a final consistent check on the acceptance of the detector system and the potential contamination, we show in Fig.~\ref{fig:slice} the dependence of $\Gamma$ and $\Gamma_0$ on $R$. 
In the present work, the mean $\Gamma$ and $\Gamma_0$ values are 0.894 and 0.992. 
A significant deviation of more than 2$\sigma$ from the weighted mean will be excluded in the final cross section determination. 

In Fig.~\ref{fig:cone} we present schematically the data analysis procedure for charge-changing cross section measurements. Basically, it is a multi-step process to filter and select only ``good" events. In $^{28}$Si case, 
42.9\% incident ions of various nuclides are eventually used for the cross section determination.

\begin{figurehere}
    \centering
    \includegraphics[width=3in]{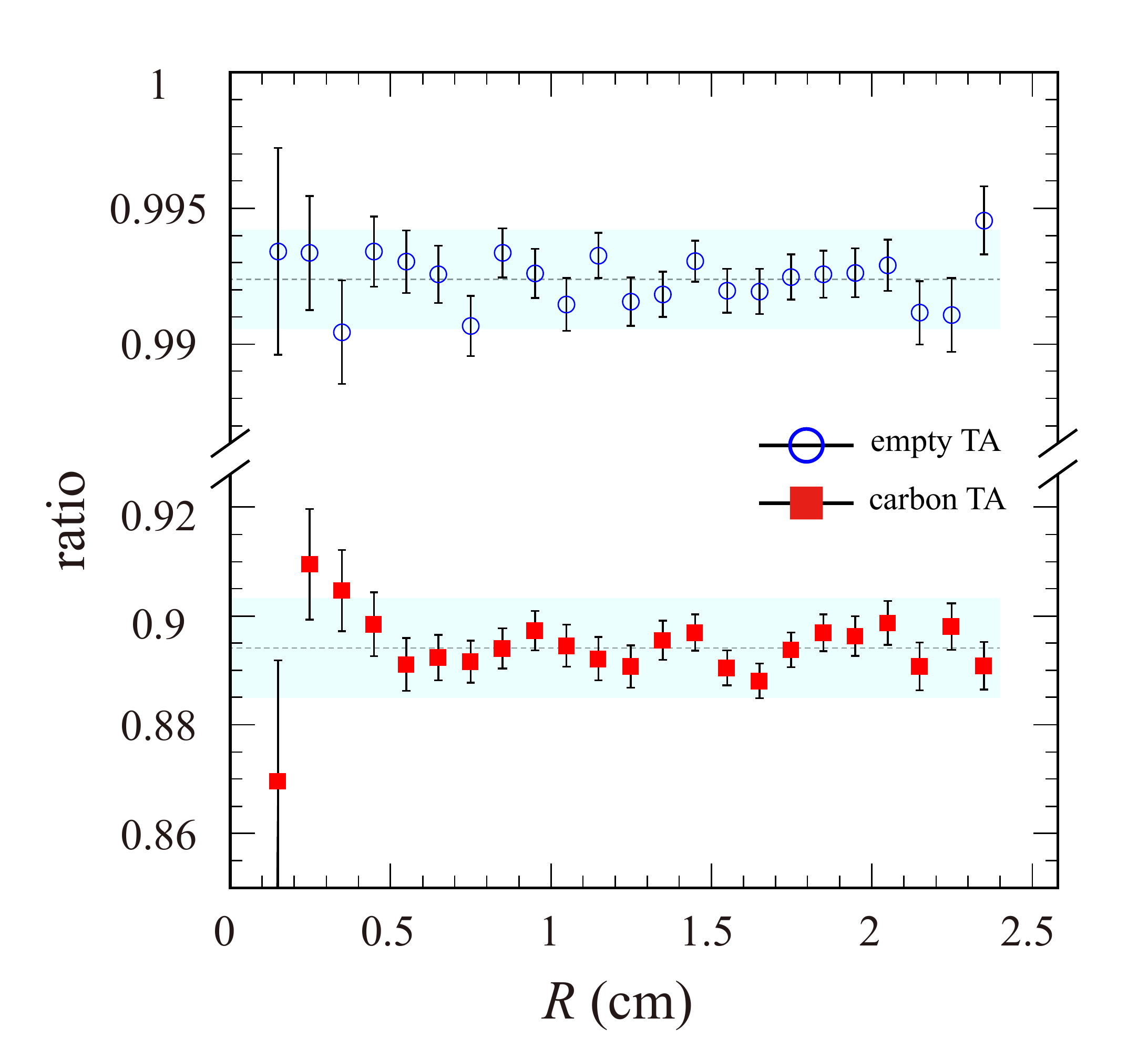}
    \figcaption{(Color online) Dependence of the \textit{$\Gamma_0$} (open symbol) and \textit{$\Gamma$} (filled symbol) ratios on $R$.
 The weighted averages and the $\pm2\sigma$ ranges are indicated for both \textit{$\Gamma_0$} and \textit{$\Gamma$}.}
    \label{fig:slice}
\end{figurehere}

\begin{figurehere}
	\centering
	\includegraphics[width=3in]{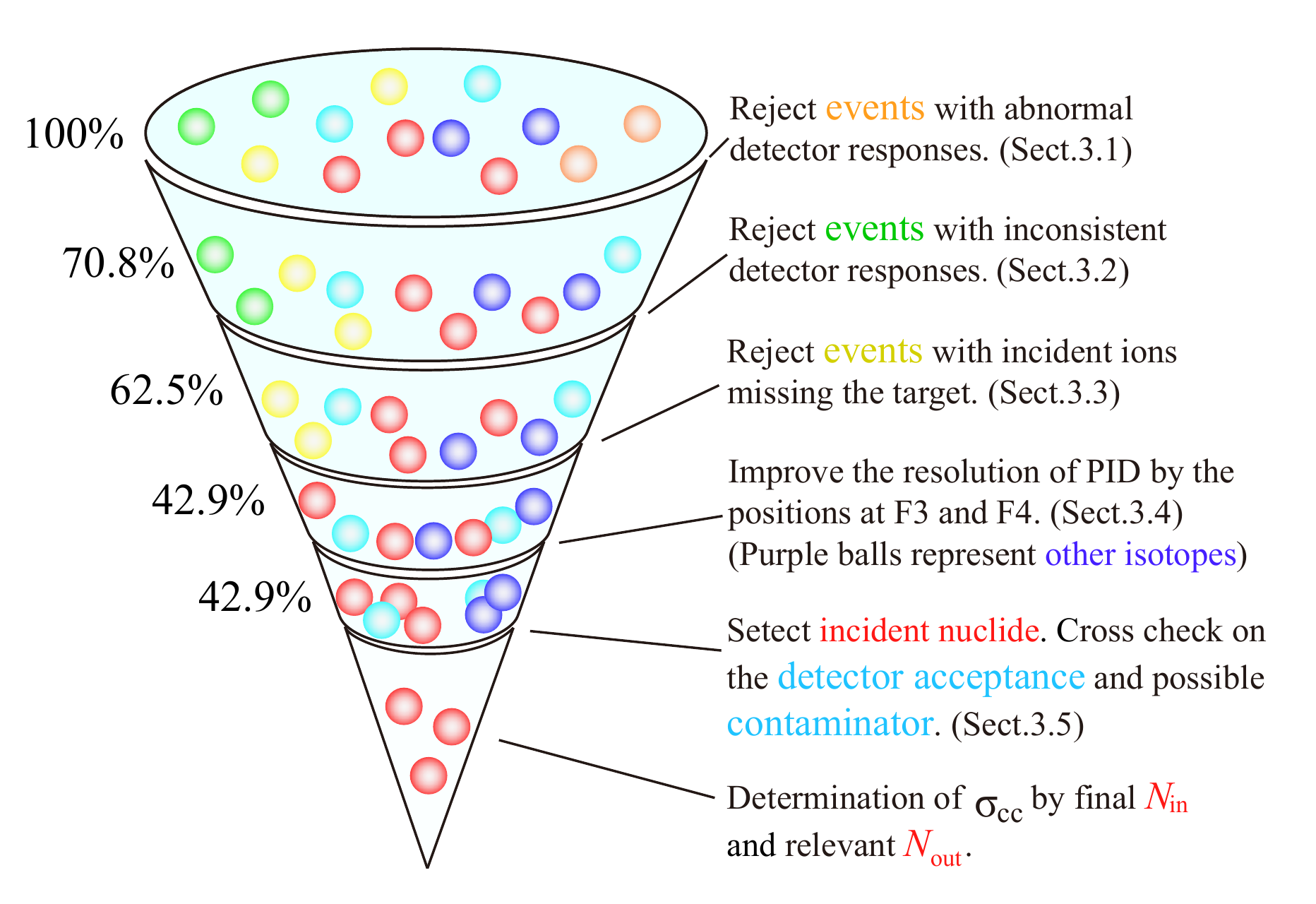}
	\figcaption{(Color online) Flow diagram of the data analysis procedure. Colored balls represent various events from the raw PID. The ratios of the number of events that survive to the total counts at each step are listed on the left. }
	\label{fig:cone}
\end{figurehere}

\section{Charge-changing cross section}\label{sec_discuss}

In the present experiment, the energy of $^{28}$Si ions in the center of the carbon target is determined to be 304(9) MeV/nucleon. The $\sigma_\text{cc}$ of $^{28}$Si on carbon is determined to be 1125(11) mb. The uncertainty is a square root of the statistic uncertainty and the systematic uncertainty. The statistical error of 10.5 mb dominates the final error. Concerning the systematic error, it is mainly due to the contamination from neighboring isotopes. To evaluate this uncertainty, we first project the $\Delta E$-TOF spectrum in  Fig.~\ref{fig:pidnin} onto the major and minor axes of the red ellipse, and then use multi-Gaussian functions to fit the projection spectra. The contamination uncertainty of $N_{\rm in}$ is calculated to be 0.6 mb.      

Previous $\sigma_{\text{cc}}$ measurements of N isotopes at 900 MeV/nucleon on carbon~\cite{Bag2019PLB} show that the cross sections can vary systematically by about 50 mb, depending on how to treat the information of the veto detector right in front of the target. This difference may be due to the back-scattered events from the carbon target.  
In the present work, such 
backscattering-like events would induce another hit in MWDC2. As described in Sect.~\ref{sec_exp}, 
our approach to determine the incoming particle's trajectory already exclude
the use of hits induced by back-scattered particles from TA, since the hit positions of backscattering are likely very different from the relevant hits of the incident particles. 
Moreover, in cases that there is more than one trajectory with SSE smaller than 0.04, we have evaluated the uncertainty in the final cross section when the real incoming particle should have the trajectory relevant to the second best fit. This uncertainty is examined to be smaller than 1 mb. We conclude that the backscattering-like events have negligible effects in the present work.

\begin{figurehere}
    \centering
    \includegraphics[width=3.5in]{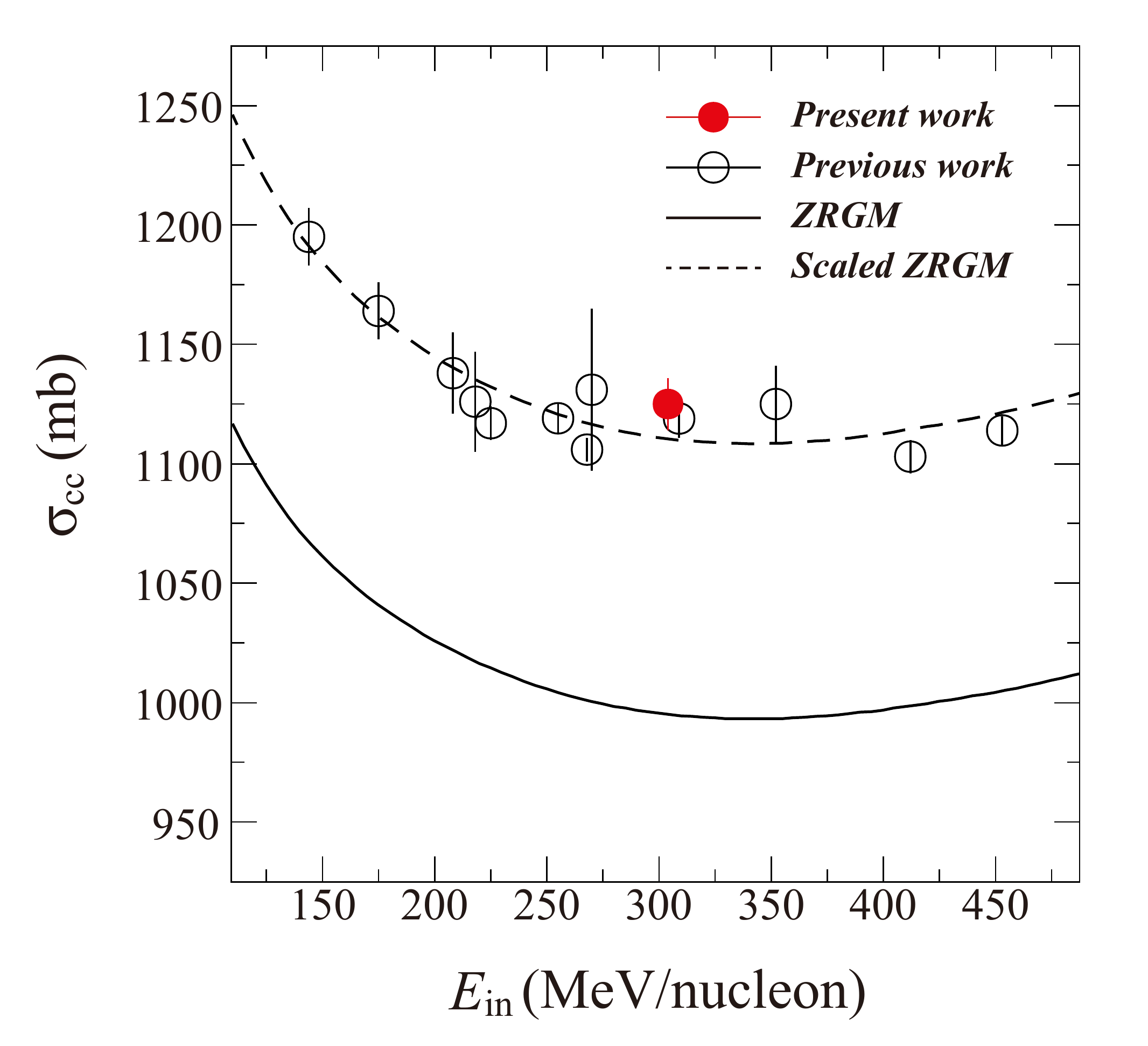}
    \figcaption{(Color online) $\sigma_{\text{cc}}$ (symbols) as a function of incident energies for $^{28}$Si+$^{12}$C. The present measurement is shown by the solid circle and the literature data represented by the open symbols are from Refs.~\cite{Liguangshuai2023PRC,Yamag2010PRC,Zeitl2007NPA,Sawah2017NPA}. ZRGM and the scaled ZRGM results are shown as the solid and dashed lines, respectively.}
    \label{fig:result}
\end{figurehere}    

In Fig.~\ref{fig:result}, we summarized the available charge-changing cross sections of $^{28}$Si on carbon at similar energies. Especially, the cross sections determined at 268, 270 and 309 MeV/nucleon were 1106(5) mb, 1131(34) mb, and 1119(8) mb, respectively~\cite{Zeitl2007NPA,Sawah2017NPA,Yamag2010PRC}. Our result is in good agreement with the previous measurements at similar energies within the uncertainties~\cite{Liguangshuai2023PRC,Zeitl2007NPA,Yamag2010PRC,Sawah2017NPA}. Regarding the energy dependence, $\sigma_{\rm cc}$ first decreases from 100 MeV/nucleon to about 200 MeV/nucleon and then remains almost constant up to 350 MeV/nucleon. 

We have performed the calculations using the zero-range optical-limit Glauber model (ZRGM)~\cite{Yamag2010PRC}. The proton density distribution of $^{28}$Si is described by the two-parameter Fermi model (2pF):
\begin{equation}
    \rho(r)=\frac{\rho_0}{1+exp(\frac{r-R}{a})}\;,~
\end{equation}
where $R$ = 3.2331 fm, $a$ = 0.4472 fm and $\rho_0$ = 0.0231 fm$^{-3}$ are the half-density radius, diffuseness and central densities, respectively. $R$ and $a$ are determined by the density normalization and the $rms$ proton radius $R_\text{p}=3.006$ fm from the experimental known charge radius~\cite{Angeli2013ADND}.
For the carbon target, we use a harmonic oscillator (HO) density distribution of $^{12}$C~\cite{Tanak2022PRC}:
\begin{equation}
    \rho(r)=\rho(0)\left[1+\frac{C-2}{3}(\frac{r}{w})^2\right]exp\left[-(\frac{r}{w})^2\right]\;,~
\end{equation}
where $C$ is the number of protons or neutrons and $w$ is the radius parameter. The experimental $R_\text{p}$ of $^{12}$C~\cite{Angeli2013ADND} and interaction cross section of 950 MeV/nucleon $^{12}$C on carbon target~\cite{Ozawa2001NPA} are available. The HO function parameters reproducing these results are adopted. The radius parameters for proton and neutron are $w_{\rm p}$ = 1.5815 fm and $w_{\rm n}$ = 1.5687 fm, respectively. Central densities $\rho_{\rm 0p}$ and $\rho_{0n}$ are  0.0908 fm$^{-3}$ and 0.0930 fm$^{-3}$, respectively.

We found that the ZRGM calculations are systematically smaller than the experimental data by about 10\%. This is consistent with the conclusion in Ref.~\cite{Yamag2010PRC}.
By multiplying the ZRGM results by a factor of 1.118, the scaling results can nicely reproduce the experimental data over a large incident energies ranging from 100 to 500~MeV/nucleon.  

\section{SUMMARY}\label{sec_sum}

In summary, we took a recent $\sigma_{\text{cc}}$ measurement of $^{28}$Si on carbon at RIBLL2 as an example to depict the data analysis procedure. The key is to use only those ``good" incident ions that bombard the target in a very well-defined position and angle. This requires all relevant detectors to have not only good but consistent responses. We also introduced a method on how to resolve the events on the target from those on the target frame. The data analysis procedure here will be used as a standard routine for the transmission measurement at RIBLL2 and the high-energy fragment separator (HFRS) at high-intensity heavy-ion accelerator facility
(HIAF)~\cite{Zhoux2022AAPPS}. Finally, we reported a new $\sigma_{\rm cc}$ result of $^{28}$Si on C target at 304(9) MeV/nucleon. It is found to be well consistent with the existing data at similar incident energies. In the energy range of 100-500 MeV/nucleon, ZRGM typically underestimates the cross sections by about 10\%.  

\section*{Acknowledgments}\label{sec_acknow}

The authors thank the HIRFL-CSR accelerator staff for offering the primary and secondary beams during the experiment. This work is supported by the National Natural Science Foundation of China (Grants No. U1832211, No. 11961141004, No. 11922501, No. 11475014, and No. 11905260), the Western Light Project of Chinese Academy of Sciences, and Anhui Provincial Natural Science Foundation (Grant No. 2008085MA17).

\end{multicols}

\vspace{-1mm}
\centerline{\rule{80mm}{0.1pt}}
\vspace{2mm}

\begin{multicols}{2}

\vspace{3mm}
\end{multicols}

\clearpage

\end{document}